\documentclass[universe,article,accept,pdftex,moreauthors]{Definitions/mdpi} 

\makeatletter
\let\c@lofdepth\relax
\let\c@lotdepth\relax
\makeatother
\usepackage{subfigure}
\makeatletter
\renewcommand{\@thesubfigure}{\normalsize(\textbf{\alph{subfigure}})}
\makeatother

\usepackage{color}
\usepackage{amsfonts}
\usepackage{mathrsfs}
\usepackage{epsfig}
\usepackage{graphicx}%
\usepackage{amsmath}

\newcommand{\blue}[1]{\textcolor{black}{{#1}}}

\newcommand\beq{\begin{equation}}
\newcommand\eeq{\end{equation}}
\newcommand\beqn{\begin{eqnarray}}
\newcommand\eeqn{\end{eqnarray}}

\firstpage{1} 
\pubvolume{1}
\issuenum{1}
\articlenumber{0}
\pubyear{2024}
\copyrightyear{2024}
\externaleditor{Academic Editor: Firstname \linebreak Lastname}
\datereceived{23 July 2024} 
\daterevised{12 August 2024} % Comment out if no revised date
\dateaccepted{13 August 2024} 
\datepublished{ } 
\hreflink{https://doi.org/} % If needed use \linebreak
\Title{{From} the Janis--Newman--Winicour Naked Singularities to the Einstein--Maxwell Phantom Wormholes}

\TitleCitation{From the Janis--Newman--Winicour Naked Singularities to the Einstein--Maxwell Phantom Wormholes}

\Author{{Changjun} %MDPI: Please carefully check the accuracy of names and affiliations.
 Gao $^{1,}${*$^{,\dagger}$} {and} %MDPI: We added front note to keep consistent with the standard note, please confirm if it should be retained.
 Jianhui Qiu $^{2,}${*$^{,\dagger}$} %MDPI: We added asterisk behind the correspondence author, according to the submitted system. Please confirm. The following highlight is the same.
}

\AuthorNames{Changjun Gao and Jianhui Qiu}

\AuthorCitation{Gao, C.; Qiu, J.}
\address{%
$^{1}$ \quad National Astronomical Observatories, Chinese Academy of Sciences, 20A Datun Road, Beijing 100101, China\\
$^{2}$ \quad School of Astronomy and Space Sciences, University of Chinese Academy of Sciences,
19A Yuquan Road, Beijing 100049, China}

\corres{Correspondence: gaocj@nao.cas.cn (C.G.); jhqiu@nao.cas.cn (J.Q.)}

\firstnote{These authors contributed equally to this work.}  % Current address should not be the same as any items in the Affiliation section.
\abstract{The Janis--Newman--Winicour spacetime corresponds to a static spherically symmetric solution of Einstein equations with the energy momentum tensor of a massless quintessence field. It is understood that the spacetime describes a naked singularity. The solution has two parameters, $b$ and $s$. To our knowledge, the exact physical meaning of the two parameters is still unclear. In this paper, starting from the Janis--Newman--Winicour naked singularity solution, we first obtain a wormhole solution by a complex transformation. Then, letting the parameter $s$ {approach} infinity, we obtain the well-known exponential wormhole solution. After that, we embed both the Janis--Newman--Winicour naked singularity and its wormhole counterpart in the background of a de Sitter or anti-de Sitter universe with the energy momentum tensor of massive quintessence and massive phantom fields, respectively. To our surprise, the resulting quintessence potential is actually the dilaton potential found by one of us. It indicates that, by modulating the parameters in the charged dilaton black hole solutions, we can obtain the Janis--Newman--Winicour solution. Furthermore, a charged wormhole solution is obtained by performing a complex transformation on the charged dilaton black hole solutions in the background of a de Sitter or anti-de Sitter universe. We eventually find that $s$ is actually related to the coupling constant of the dilaton field to the Maxwell field and $b$ is related to a negative mass for the dilaton black holes. A negative black hole mass is physically forbidden. Therefore, we conclude that the Janis--Newman--Winicour naked singularity solution is not {physically allowed.}}


\begin{document}

%%%%%%%%%%%%%%%%%%%%%%%%%%%%%%%%%%%%%%%%%%
\section{Introduction}\label{sec1}
The Janis--Newman--Winicour (JNW) naked singularity solution describes the most general, static, spherically symmetric, and~asymptotically spatially flat spacetime in the Einstein-massless quintessence systems~\cite{jnw:1968}. The~solution was first discovered by Fisher~\cite{fisher:1948} in 1948. It was then rediscovered by Janis, Newman, and Winicour in an isotropic coordinate system in 1968. Thereafter, Wyman~\cite{wyman:1981} discovered once again the solution in the Schwarzschild coordinate system. Later, the~equivalence of the Wyman solution with the JNW spacetime was proved by Virbhadra~\cite{virb:1997} in 1997. Agnese and Camera~\cite{agn:1985}
rewrite the Wyman solution in a more compact expression. Roberts~\cite{rob:1993} proved that the most general static spherically symmetric solution to the Einstein equations with the source of massless scalar field is asymptotically flat and this is exactly the Wyman solution. 
Bronnikov and Khodunov~\cite{bron:1979} studied the stability problem of JNW spacetime. Chew and Lim utilize a symmetric scalar potential to make the JNW spacetime regular, and a gravitational soliton solution is numerically constructed~\cite{chew:2024}. In~short, a~lot of work on JNW spacetime has been conducted in various research fields, such as gravitational lensing and relativistic images~\cite{gyu:2008,virb:2002,virb:1998,virb:2008,gyu:2019}, black hole accretion and shadows~\cite{yang:2015,tak:2004,chow:2012,sub:2020,pal:2023}, and so on~\cite{sta:1,sta:2,sta:3,matos:1,matos:2,matos:3,matos:4,matos:5,matos:6,matos:7,matos:8}.

The important claim here is that, actually, Sadhu and Suneeta~\cite{sad:2013} proposed that the JNW spacetime can be obtained from a class of charged dilaton black hole solutions~\cite{gib:1988,gar:1991}. In~practice, they argue that the JNW spacetime is obtained from the charged dilaton black hole solutions in~\cite{gar:1991} by setting $r_{+}=0$ but with $r_{-}\neq{0}$ in Equations~\eqref{eq20}--\eqref{eq23} of that paper. However, we now point out that one cannot set $r_{+}=0$ naively because $r_+$ has a definite physical meaning and it is always positive. If~one arbitrarily lets $r_+=0$, that means we must assume that the black hole has a negative mass. Then, it is not a surprise that we shall obtain a naked singularity. This is just like if we merely mathematically, while not physically, assume that the mass parameter $M$ in the Schwarzschild metric is negative:

\begin{equation}
ds^2=-\left(1-\frac{2M}{r}\right)dt^2+\left(1-\frac{2M}{r}\right)^{-1}dr^2+r^2d\Omega^2\;.
\end{equation} 
Then, the well-known Schwarzschild singularity becomes naked to us. The~reason that the JNW solution brings us a naked singularity is that we have assumed a negative mass in advance. Therefore, from~the point of view of dilaton black hole solutions, the~JNW naked singularity solution is unphysical. As is known, the~JNW spacetime is asymptotically flat in space. In~other words, the~naked singularity is embedded in the background of Minkowski spacetime. Recognizing this, the initial motivation of this work is in fact to embed the naked singularity in the background of a de Sitter or anti-de Sitter universe (DAU). However, we finally realize that the JNW solution is~unphysical. 

The paper is organized as follows. In~Section \ref{sec2}, we rewrite the JNW solution in the Schwarzschild coordinate system in a desirable expression. The~initial parameter $b$ is replaced by $2M/s$, with $M$ standing for the mass of the singularity. In~Section \ref{sec3}, by~using a complex transformation method on the quintessence field for the JNW solution, we achieve an exact wormhole solution. The~reason for the naked singularity transforming into a wormhole is that the quintessence field becomes a phantom field~\cite{caldwell:2002}. In~Section \ref{sec4}, we let the parameter $s$ approach positive infinity, and then the well-known exponential wormhole solution is produced~\cite{pap:1954}. In~Section \ref{sec5}, we embed the naked singularity and wormhole in the background of a de Sitter universe in an isotropic coordinate system.  In~Section \ref{sec6}, via coordinate transformations from isotropic coordinates to Schwarzschild coordinates, we construct the naked singularity or wormhole in the background of DAU. We find that the corresponding spacetime structure is very interesting. For~example, there are both black holes and wormholes in the same spacetime. In~Sections \ref{sec7} and \ref{sec8}, in~the framework of a massive quintessence field and phantom field, we derive their scalar potentials with respect  to the metric. To~our surprise, the~resulting quintessence potential is exactly the dilaton potential derived by one of us~\cite{gao:2004}. We shall show this point in Section \ref{sec9}. Subsequently, this motivates us to investigate whether there exists a relation between the JNW solution and the dilaton black hole solution, just as proposed by Sadhu and Suneeta~\cite{sad:2013}. {In~Section \ref{sec10},} we are inspired to seek for the charged wormhole solutions from the charged dilaton counterpart. During~this process, we realize that the JNW naked singularity solution is unphysical because nature does not allow the dilaton black hole to have a negative mass. Section \ref{sec11} gives the conclusion and discussion. Throughout the paper, we adopt the system of units in which $G=c=\hbar=1$ and the metric signature
$(-, +, +, +)$.

\section{A Useful Expression for the JNW~Metric}\label{sec2}
When a massless quintessence ﬁeld is minimally coupled to gravity, the~associated Einstein equations are given by
\begin{equation}\label{8pi}
R_{\mu\nu}=-\nabla_{\mu}\phi\nabla_{\nu}\phi\;,
\end{equation}
where $R_{\mu\nu}$ is the Ricci tensor. Subsequently,  the~equation of motion for the quintessence field is
\begin{equation}
\nabla^2\phi=0\;,
\end{equation}
where $\nabla^2$ is the four-dimensional Laplace operator.  Janis, Newman, and Winicour\mbox{ (JNW) \cite{jnw:1968}} have shown that  the metric
\begin{eqnarray}\label{JNW-Q1}
dl^2&=&-\left(1-\frac{b}{r}\right)^sdt^2+\left(1-\frac{b}{r}\right)^{-s}dr^2+r^2\left(1-\frac{b}{r}\right)^{1-s}d\Omega^2\;,
\end{eqnarray}
with
\begin{eqnarray}
d\Omega^2&=&d\theta^2+\sin^2\theta d\varphi^2\;,
\end{eqnarray}
and the field
\begin{eqnarray}
\phi&=&\frac{\sqrt{2}}{2}\sqrt{1-s^2}\ln\left(1-\frac{b}{r}\right)\;,
\end{eqnarray}
satisfy the Einstein equations and the equation of motion for the quintessence field. The~solution is popularly known as the JNW spacetime in the literature. Here, $r$ represents the radial coordinate and $s$ is a dimensionless integration constant.  $s$  runs over the range of
\begin{eqnarray}
0\leq s\leq 1\;.
\end{eqnarray}

In particular, when $s=1$, the~scalar field vanishes and the Schwarzschild solution is recovered. In~general, we should require that
\begin{eqnarray}
r\geq b\;,
\end{eqnarray}
 in order to obtain a physical spacetime.  The~physical significance of $b$ and $s$ is unclear. The~Ricci scalar of the spacetime is
\begin{eqnarray}\label{ricci}
R=\frac{b^2\left(s^2-1\right)}{2r^{2+s}\left(r-b\right)^{2-s}}\;.
\end{eqnarray}

It reveals that when $s\neq 1$, there is a curvature singularity at $r=b$ provided that $b>0$. Since the  singularity is not cloaked by an event horizon, this metric represents a naked singularity. Therefore, we conﬁne ourselves in the region $r>b$. By~expanding the metric in the order of $b/r$ , we find that the parameter $b$ is related to the physical mass of the spacetime $M$ by
\begin{eqnarray}\label{mass}
b&=&\frac{{2M}}{s}\;.
\end{eqnarray}

When $s=1$, it is exactly the Schwarzschild radius. Then, the JNW solution becomes the following expression:
\begin{eqnarray}\label{JNW-Q}
dl^2&=&-\left(1-\frac{2M}{sr}\right)^sdt^2+\left(1-\frac{2M}{sr}\right)^{-s}dr^2+r^2\left(1-\frac{2M}{sr}\right)^{1-s}d\Omega^2\;,
\end{eqnarray}
and the field is
\begin{eqnarray}
\phi&=&\frac{\sqrt{2}}{2}\sqrt{1-s^2}\ln\left(1-\frac{2M}{sr}\right)\;.
\end{eqnarray}

In the following sections, we shall see how the above parameterized expression for the metric is useful to~us. 

\clearpage  
\section{A New Traversable~Wormhole}\label{sec3}
\subsection{The~Solution}
\textls[-15]{In this subsection, we show how we will obtain a new traversable wormhole  by starting from the JNW spacetime.  To~illustrate this point, we make the following \mbox{complex transformation:}}
\begin{equation}\label{ct}
\phi\longrightarrow e^{i\frac{\pi}{2}}\phi\;,
\end{equation}
where $i$ is the {imaginary } unit. Then, the Einstein equations turn out to be
\begin{equation}
R_{\mu\nu}=\nabla_{\mu}\phi\nabla_{\nu}\phi\;,
\end{equation}
and the equation of motion for the massless scalar field is invariant
\begin{equation}
\nabla^2\phi=0\;.
\end{equation}

Now, the scalar field  appears as a phantom field.   We find the metric
\begin{eqnarray}\label{JNW-Phantom}
dl^2&=&-\left(1-\frac{2M}{sr}\right)^sdt^2+\left(1-\frac{2M}{sr}\right)^{-s}dr^2+r^2\left(1-\frac{2M}{sr}\right)^{1-s}d\Omega^2\;,
\end{eqnarray}
 and the field
\begin{eqnarray}\label{phantom}
\phi&=&\frac{\sqrt{2}}{2}\sqrt{s^2-1}\ln\left(1-\frac{2M}{sr}\right)\;,
\end{eqnarray}
that solve both the Einstein equations and the phantom field equation.  The~argument under the root-sign is always non-negative. Therefore,  we should now require
\begin{eqnarray}
s>1\;.
\end{eqnarray}

We see that the expression of the metric is the same as JNW spacetime. However, we shall find sooner that it denotes not a naked singularity but a wormhole, provided that $s>2$. As~indicated by the Ricci scalar  and other curvature invariants (for example, the~Kretschmann quadratic invariant),  when
\begin{eqnarray}
s\geq{2}\;,
\end{eqnarray}
the sphere $r=b$ or $r=2M/s$ is no longer a curvature singularity.  We note that the method of complex transformation, i.e.,~Equation~(\ref{ct}), is attractive because one can obtain new solutions from the known ones with the source of quintessence fields. For~example, we would most likely find new phantom hairy black hole solutions from the known quintessence ones. In~practice, Nozawa1 and Torii have developed a considerable family of exact solutions to the Einstein phantom theories \cite {noz:2023} by using this~method.

In order to show that when $s>2$, it leads to a wormhole, we consider the area of the spherical surfaces of constant $r$ coordinate following Ref.~{\cite{boo:2018}}:
\begin{eqnarray}\label{eq20}
A=4\pi r^2\left(1-\frac{2M}{sr}\right)^{1-s}\;.
\end{eqnarray}

Then, we find
\begin{eqnarray}\label{eq21}
\frac{dA}{dr}=8\pi r^s\left(s r-2M\right)^{-s} s^{s-1}\left[sr-\left(1+s\right)M\right]\;,
\end{eqnarray}
\begin{eqnarray}\label{eq22}
\frac{d^2A}{dr^2}&=&8\pi r^{s-1}\left(sr-2M\right)^{-s-1} s^s\nonumber\\&&\left(sr^2-2rM-2srM+2M^2+2M^2s\right)\;.
\end{eqnarray}

Noting that the domain for the $r$-coordinate is $r\in\left(\frac{2M}{s}, +\infty\right)$, we see that we \mbox{always have}
\begin{eqnarray}\label{eq23}
\frac{d^2A}{dr^2}>0\;,
\end{eqnarray}
in this domain, provided that $s>2$.  Namely, the~area is a concave function of the $r$ coordinate.  On~the other hand, the~equation $\frac{dA}{dr}=0$ gives  the throat (minimum) at
\begin{eqnarray}
r_{T}=\frac{\left(1+s\right)M}{s}, 
\end{eqnarray}
where it satisﬁes the “ﬂare out” condition $A^{''}_{r=r_{T}}>0$. We observe that all the metric components are ﬁnite at the throat. So it is now enough to guarantee that the surface $r=r_{T}$ is a traversable wormhole following the definition of Morris and Thorne~\cite{mor:1988,mor:1988b,vis:1989,vis:1989b}. The~geometry described by the wormhole metric clearly has no horizons, since when $r\in\left(\frac{2M}{s}, +\infty\right)$, we have
\begin{eqnarray}
g_{00}\neq{0}\;.
\end{eqnarray}

Here, $(0,1,2,3)$ denote $(t,r,\theta,\phi)$.  As~already demonstrated, there is a
traversable wormhole throat located at $r=r_{min}$, where the area of the spherical surfaces is minimized, and~the “ﬂare out” condition is satisﬁed. All of the curvature components and the invariant Kretschmann scalar are ﬁnite everywhere in the wormhole spacetime. As~an example, we consider $s=3$. We provide a coordinate transformation as follows:
\begin{eqnarray}
r=\frac{1}{2}x+\frac{1}{6}\sqrt{9x^2-24xM}\;.
\end{eqnarray}

The metric is
\begin{eqnarray}
&&ds^2=-\left[1-\frac{4M}{3x+\sqrt{9x^2-24xM}}\right]^3dt^2+\frac{\left(\frac{1}{2}+\frac{3x-4M}{2\sqrt{9x^2-24xM}}\right)^2}{\left[1-\frac{4M}{3x+\sqrt{9x^2-24xM}}\right]^3}dx^2+x^2d\Omega^2\;.
\end{eqnarray}

The domain for the $x$-coordinate is $x\in\left[\frac{8M}{3}, +\infty\right)$; we see that  $x_{min}=\frac{8M}{3}$ corresponds to the throat of the wormhole. In~all, when $s>2$, we obtain a traversable wormhole with the throat at $r=\frac{M\left(s+1\right)}{s}$.

\begin{figure}[H]
	%\centering
	\includegraphics[width=8cm,height=6cm]{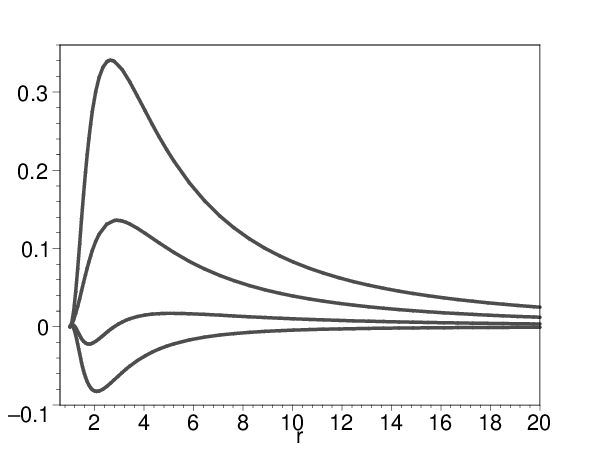}
	\caption{The effective potential of odd parity $V_{eff,odd}$ with the radius $r$ assuming $M=1\;, s=2$ for four different cases $l=
0, 1, 2, 3$, from~down to up, respectively. The potential is asymptotically vanishing at both the wormhole throat and infinity. } 
	\label{oddpotential.eps}
\end{figure}

 {\subsection{Stability~Analysis}}
{ We have proven that the wormhole solution is traversable by considering the flaring-out condition near the throat. In~this subsection, we address the stability problem by firstly studying the adiabatic sound speed~\cite{tak:2023}. The~square of adiabatic sound speed is defined by} 
\vspace{-6pt}
\blue{\begin{eqnarray}
&&v_s^2=\frac{\delta\left\langle{p}\right\rangle}{\delta\rho}\;,
\end{eqnarray}
\vspace{-6pt}
where $\left\langle{p}\right\rangle$ represents the average pressure across the three spatial dimensions, namely $\left\langle{p}\right\rangle=({p_r+2p_t})/{3}$. The~energy density $\rho$, and the~radial and the tangential pressures, $p_r$ and $p_t$, are deduced from Equation~(\ref{JNW-Phantom}) as follows:
\begin{eqnarray}
&&\rho=p_r=-p_t=\frac{M^2\left(1-s^2\right)}{8{\pi}r^4s^2}\left(1-\frac{2M}{sr}\right)^{s-2}\;.
\end{eqnarray}
{Thus}, %MDPI: %MDPI: Please confirm if no-indent paragraph should be retained. The following highlights are the same.
 we have the radial squared sound speed $v_{sr}^2$, the~tangential squared sound speed $v_{st}$, and the average squared of sound speed:
\begin{eqnarray}
v_{sr}^2=\frac{\delta{p_r}}{\delta\rho}=1\;,\ \ \ v_{st}^2=\frac{\delta{p_t}}{\delta\rho}=-1<0\;,\ \ \ v_s^2=\frac{\delta\left\langle{p}\right\rangle}{\delta\rho}=-\frac{1}{3}<0\;.
\end{eqnarray}
\textls[-25]{{In this sense}, the~solution is unstable because of the negative of the sound speeds. However, from~the perspective of metric perturbations, the~solution is stable. Actually, \mbox{Kobayashi et al.~\cite{kob:2012}} have explored the metric perturbations to static spherically symmetric spacetime for odd parity in the Horndeski theory, which covers the Einstein massless scalar system studied in this paper. Following their conventions,  we find}
\begin{eqnarray}
\mathcal{F}=1\;,\ \ \mathcal{G}=1\;,\ \ \mathcal{H}=1\;,
\end{eqnarray}
\textls[10]{for the Einstein massless scalar system.  
In order to avoid gradient instability,  Kobayashi et al.~\cite{kob:2012} show that}
\begin{eqnarray}
\mathcal{F}>0\;.
\end{eqnarray}
{On the other hand}, in~order to avoid the presence of ghost, they show
\begin{eqnarray}
\mathcal{G}>0\;.
\end{eqnarray}
{It is obvious} that the two conditions are satisfied for the Einstein massless system. The~squared speeds of gravitational waves along the radial direction,
$c_r^2$, and the tangential direction, $c_{t}^2$, are found to be}
\begin{eqnarray}
c_{r}^2=\frac{\mathcal{G}}{\mathcal{F}}=1\;,\ \ \ c_{t}^2=\frac{\mathcal{G}}{\mathcal{H}}=1\;.
\end{eqnarray}
{They are exactly} the square of the speed of light. The~above conditions on  $\mathcal{F},\mathcal{G}$, and $\mathcal{H}$ are necessary for the stability of wormholes. The~sufficient condition for the stability of wormholes  is that the effective potential $V_{eff,odd}$ \cite{gao:2022} satisfies the condition
\begin{eqnarray}\label{oddpot}
V_{eff,odd}=\frac{U}{f^2}\left[l\left(l+1\right)-2\right]+fU\partial_{r}\left(U\partial_{r}\frac{1}{f}\right)\geq{0}\;,
\end{eqnarray}
with
\begin{eqnarray}
U=\left(1-\frac{2M}{sr}\right)^s\;,\ \ \ \ \ f=r\left(1-\frac{2M}{sr}\right)^{\frac{1-s}{2}}\;,
\end{eqnarray}
outside the wormhole throat. We find that it is indeed the case when $l\geq 2$ and $s\geq{2}$. Therefore, in~this sense, the~wormhole is stable to metric perturbations except for $l=0, 1$.  In~Figure~\ref{oddpotential.eps}, we plot the effective potential $V_{eff,odd}$ in terms of $r$. The~potential is positive everywhere and tends to zero at both  infinity and the throat when $l\geq{2}$. As~a comparison, a~potential well appears for $l=0$ and $l=0$, which implies the instability of wormholes to metric perturbations in this~situation. 

\vspace{-6pt}

 \section{The Exponential Wormhole~Spacetime}\label{sec4}
 In this section, we show that the well-known exponential wormhole spacetime turns out to be a special case of the above wormhole solution. To~show this point, let $s$ approach positive infinity, and then we obtain, by putting $s\rightarrow+\infty$ in Equation~(\ref{JNW-Phantom}),
\begin{eqnarray}\label{JNW-P}
dl^2&=&-e^{-\frac{2M}{r}}dt^2+e^{\frac{2M}{r}}dr^2+r^2e^{\frac{2M}{r}}d\Omega^2\;.
\end{eqnarray}

In the same way, by~putting $s\rightarrow+\infty$ in Equation~(\ref{phantom}), we obtain
\begin{eqnarray}\label{JNW-Pa}
\phi&=&-\frac{{\sqrt{2}}M}{r}\;.
\end{eqnarray}

 We have checked that the above metric and the phantom  field  do solve the Einstein equations and the scalar field equation. The~metric is novel because it describes  a non-singular spacetime.  Specifically, the~Ricci scalar and the Kretschmann invariant are all finite. The~metric is known as Yilmaz  “exponential metric”  in the literature~\cite{yil:1958,yil:1971,yil:1973,cla:1973}. But~we point out that it appeared for the first time in Ref.~\cite{pap:1954} by Papapetrou. The~metric has attracted wide attention in the literature~\cite{1,2,3,4,5,6,7,8,9,10,11,12,13,14,15,sim:2021} ever since it was found. However, to~our knowledge, the~energy momentum tensor for this solution has never been found. Here, we find a suitable one.  Namely, it is sourced by a massless phantom field.  Very recently, Boonserm~et~al.~\cite{boo:2018} showed that the Yilmaz exponential metric represents a traversable wormhole. 
{We point out that the traversable wormholes have been studied in many aspects, such as the stability analysis of wormholes~\cite{1a,2a,3a}, the~resolution to
the horizon problem in cosmology~\cite{4a,5a}, the~wormhole solutions in the modified  gravities~\cite{6a,7a,8a,9a,10a,11a,12a,13a,14a,15a}, and so on~\cite{16a,17a,18a,19a}.}

\section{Naked Singularity and Wormhole in de Sitter~Universe}\label{sec5}

We have seen both the JNW (with naked singularity) solution and the wormhole solution described by Equation~(\ref{JNW-Phantom}). The~only difference is that $0<s<1$ for the former and $s>1$ for the latter. The~metric is static and asymptotically flat in space. In~this section, we construct its counterpart in the de Sitter universe. To~this end, we transform them from the Schwarzschild coordinate system to an isotropic coordinate system. So we let
\begin{equation}
r=x\left(1+\frac{M}{2sx}\right)^2\;.
\end{equation}

Then, we have
\begin{eqnarray}
dl^2=-\frac{\left(1-\frac{M}{2sx}\right)^{2s}}{\left(1+\frac{M}{2sx}\right)^{2s}}dt^2+\frac{\left(1+\frac{M}{2sx}\right)^{2s+2}}{\left(1-\frac{M}{2sx}\right)^{2s-2}}\left(dx^2+x^2 d\Omega^2\right)\;.
\end{eqnarray}

The metric is now in the isotropic coordinate system. Once the metric is written in the isotropic coordinate system, we are ready to reach the counterpart in the de Sitter universe. Following the method developed by us~\cite{gao:2004},  we obtain
\begin{eqnarray}\label{ds}
dl^2&=&-\frac{\left(1-\frac{M}{2sax}\right)^{2s}}{\left(1+\frac{M}{2sax}\right)^{2s}}dt^2+a^2\frac{\left(1+\frac{M}{2sax}\right)^{2s+2}}{\left(1-\frac{M}{2sax}\right)^{2s-2}}\left(dx^2+x^2 d\Omega^2\right)\;,
\end{eqnarray}
where
\begin{eqnarray}
a=e^{Ht}\;,
\end{eqnarray}
is the scale factor of the universe. $H$ is the Hubble constant. When $M=0$, we obtain the de Sitter universe. When $H=0$, we obtain the JNW or wormhole solution. Therefore, it represents a naked singularity or a wormhole in the de Sitter universe.  We shall see in Section 7 that the metric does satisfy the massive scalar field equation and the Einstein~equations.  

Letting $s\longrightarrow+\infty$ in Equation~(\ref{ds}), we obtain
\begin{eqnarray}\label{wormds}
dl^2=-e^{-\frac{2M}{ax}}dt^2+a^2e^{\frac{2M}{ax}}\left(dx^2+x^2 d\Omega^2\right)\;.
\end{eqnarray}

It describes  an exponential wormhole in the background of an exponential expanding de Sitter~universe. 

Does the throat of the wormhole expand with the universe? The answer is no. The~reasons are as follows. Consider the physical area of a spherical sphere $A$, which is given by
\begin{eqnarray}
A=4\pi a^2x^2\frac{\left(1+\frac{M}{2sax}\right)^{2s+2}}{\left(1-\frac{M}{2sax}\right)^{2s-2}}\;.
\end{eqnarray}

The throat of the wormhole is determined by $dA/dx=0$ for any scale factor $a$. So, the~throat is located at
\begin{eqnarray}\label{coleng}
x_T=\frac{M}{2as}\left(s+\sqrt{s^2-1}\right)\;.
\end{eqnarray}
{We emphasize} that it is a coordinate length, not a physical length. The~physical length of the radius of the throat is
\begin{eqnarray}
l_T=ax_T=\frac{M}{2s}\left(s+\sqrt{s^2-1}\right)\;.
\end{eqnarray}
{It is obviously }a constant, not growing with the expansion of the universe.  Substituting Equation~(\ref{coleng})  into the formula  of physical area, we find that the physical area of the wormhole throat is
\begin{eqnarray}\label{eq47}
A_T=\frac{\pi M^2\left(s+\sqrt{s^2-1}+1\right)^{2s+2}}{\left(s+\sqrt{s^2-1}-1\right)^{2s-2}\left(s+\sqrt{s^2-1}\right)^2s^2}\;.
\end{eqnarray}

It tells us that the physical area of the wormhole throat is a constant. It does not expand with the expansion of the~universe.

\section{Naked Singularity and Wormhole in~DAU}\label{sec6}
\unskip
\subsection{Solutions in~DAU}
\textls[-10]{In this section, we look for  the naked singularity and wormhole in an anti-de Sitter universe. To~achieve this, let us start from their counterpart in the de Sitter universe, Equation~(\ref{ds}), by~making a coordinate transformation as follows:}
\begin{eqnarray}
y=ax\;.
\end{eqnarray}

Then, we obtain
\begin{eqnarray}
dl^2&=&-\frac{\left(1-\frac{M}{sy}\right)^{2s}}{\left(1+\frac{M}{sy}\right)^{2s}}dt^2+\frac{\left(1+\frac{M}{sy}\right)^{2s+2}}{\left(1-\frac{M}{sy}\right)^{2s-2}}\cdot\left[\left(dy-Hydt\right)^2+y^2 d\Omega^2\right]\;.
\end{eqnarray}

By letting
\begin{eqnarray}
y=z-\frac{M}{s}+\sqrt{z^2-\frac{2M}{s}z}\;,
\end{eqnarray}
we obtain
\vspace{-6pt}
\begin{adjustwidth}{-\extralength}{0cm}
\begin{eqnarray}
dl^2&=&-\left[\left(1-\frac{2M}{sz}\right)^{s}-4H^2z^2{\left(1-\frac{2M}{sz}\right)^{1-s}}\right]dt^2-8Hz^2{\left(1-\frac{2M}{sz}\right)^{1-s}}dtd\ln{y}\nonumber\\&&+4z^2{\left(1-\frac{2M}{sz}\right)^{1-s}}\left[\left(d\ln{y}\right)^2+d\Omega^2\right]\;.
\end{eqnarray}
\end{adjustwidth}

Rescaling $t$, $H$, and $l$ as follows:
\begin{eqnarray}
t\rightarrow 2t\;,\ \  H\rightarrow \frac{H}{2}\;,\ \  l\rightarrow 2l\;.
\end{eqnarray}
we have
\vspace{-6pt}
\begin{adjustwidth}{-\extralength}{0cm}
\begin{eqnarray}
dl^2&=&-\left[\left(1-\frac{2M}{sz}\right)^{s}-H^2z^2{\left(1-\frac{2M}{sz}\right)^{1-s}}\right]dt^2-2Hz^2{\left(1-\frac{2M}{sz}\right)^{1-s}}dtd\ln{y}\nonumber\\&&+z^2{\left(1-\frac{2M}{sz}\right)^{1-s}}\left[\left(d\ln{y}\right)^2+d\Omega^2\right]\;.
\end{eqnarray}
\end{adjustwidth}

Introducing a new time coordinate $T$ and setting
\begin{eqnarray}
dt&=&dT-{\frac{Hz^2\left(1-\frac{2M}{sz}\right)^{1-s}d\ln{y}}{\left(1-\frac{2M}{sz}\right)^{s}-H^2z^2{\left(1-\frac{2M}{sz}\right)^{1-s}}}}\;,
\end{eqnarray}
 we~obtain
\vspace{-6pt}
\begin{adjustwidth}{-\extralength}{0cm}
%\centering %% If there is a figure in wide page, please release command \centering
\begin{eqnarray}
dl^2&=&-\left[\left(1-\frac{2M}{sz}\right)^s-H^2z^2\left(1-\frac{2M}{sz}\right)^{1-s}\right]dT^2+\left[\left(1-\frac{2M}{sz}\right)^s-H^2z^2\left(1-\frac{2M}{sz}\right)^{1-s}\right]^{-1}dz^2\nonumber\\&&+z^2\left(1-\frac{2M}{sz}\right)^{1-s}d\Omega^2\;.
\end{eqnarray}
\end{adjustwidth}

Now, the metric functions are only dependent on the radial coordinate $z$. That is to say, we obtain a static spacetime.  Replacing  $H^2$  by $\Lambda/3$ with $\Lambda$, the cosmological constant,  we~obtain
\vspace{-6pt}
\begin{adjustwidth}{-\extralength}{0cm}
%\centering %% If there is a figure in wide page, please release command \centering
\begin{eqnarray}\label{desire}
dl^2&=&-\left[\left(1-\frac{2M}{sz}\right)^s-\frac{1}{3}\Lambda z^2\left(1-\frac{2M}{sz}\right)^{1-s}\right]dT^2+\left[\left(1-\frac{2M}{sz}\right)^s-\frac{1}{3}\Lambda z^2\left(1-\frac{2M}{sz}\right)^{1-s}\right]^{-1}dz^2\nonumber\\&&+z^2\left(1-\frac{2M}{sz}\right)^{1-s}d\Omega^2\;,
\end{eqnarray}
\end{adjustwidth}
i.e., the naked singularity for $s<1$, the~Schwarzschild black hole for $s=1$, and the wormhole for $s>1$ in DAU. When $\Lambda=0$, we recover the JNW spacetime or wormhole spacetime. When $\Lambda>0$, it is for the de Sitter universe and $\Lambda<0$ for the anti-de Sitter universe.  On~the other hand,  when $s=1$, we recover  the Schwarzschild--de Sitter (or anti-de Sitter) spacetime.  
Finally,  when
\begin{eqnarray}
s\longrightarrow+\infty\;,
\end{eqnarray}
we have
\begin{eqnarray}\label{lambda}
dl^2&=&-\left(e^{-\frac{2M}{z}}-\frac{1}{3}\Lambda z^2 e^{\frac{2M}{z}}\right)dT^2+\left(e^{-\frac{2M}{z}}-\frac{1}{3}\Lambda z^2 e^{\frac{2M}{z}}\right)^{-1}dz^2+z^2e^{\frac{2M}{z}}d\Omega^2\;.
\end{eqnarray}

When $M=0$, we obtain the de Sitter or ani-de Sitter spacetime. When $\Lambda=0$, we have the exponential wormhole spacetime. So it is the static form for an exponential wormhole in de Sitter or anti-de Sitter spacetime. In~the background of anti-de Sitter spacetime, there are no horizons except for the throat of the wormhole at $r=M$. However, in~the background of de Sitter spacetime, the~structure of spacetime is~nontrivial.

\subsection{Spacetime Structure of JNW in~DAU}
When $s=1$, we have the Schwarzschild--de Sitter (or anti-de Sitter) solution. As is well known, there are generally two horizons, the~black hole horizon and the cosmic horizon, together with a physical singularity in the Schwarzschild--de Sitter spacetime. When $0<s<1$, we have the JNW--(anti-)de Sitter solution.  Like the Schwarzschild--de Sitter solution, there are also  generally two horizons, the~black hole horizon $z_b$ and the~cosmic horizon $z_c$ ($z_{b,c}$ are determined by $g_{00}(z=z_{b,c})=0$)), and a physical singularity $z_s=\frac{2M}{s}$ in this spacetime. When the black hole horizon  and the cosmic  horizon coincide, we have
\begin{eqnarray}
g_{00}=0\;,\ \ \ \ \ \  \frac{d{g_{00}}}{dz}=0\;.
\end{eqnarray}

They lead to a critical value $\Lambda_{c}$ for $\Lambda$:
\begin{eqnarray}
\Lambda_{c}=\frac{3s^2\left(2s-1\right)^{2s-1}}{M^2\left(2s+1\right)^{2s+1}}\;,
\end{eqnarray}

It is obvious that $s=\frac{1}{2}$ is a special value. So we conduct a discussion  in three different situations,  i.e.,~$\frac{1}{2}<s<1$, $s=\frac{1}{2}$, and  $0<s<\frac{1}{2}$,

\subsubsection{$\frac{1}{2}<s<1$}
In this case, we have the following {conclusion. }%MDPI: Please confirm if this bold should be retained and keep all symbols uniform through the whole paper including equations and text.

$\boldsymbol{\alpha}$.
 {When} %MDPI: Please check if these highlighted paragraph should be list format.
\begin{eqnarray}
\Lambda\leq 0\;,
\end{eqnarray}
we always have $g_{00}<0$ and $g_{11}>0$. 
Therefore, there are no horizons in this spacetime and the singularity $z_s=\frac{2M}{s}$ is~naked.

$\boldsymbol{\beta}$. {When}
\begin{eqnarray}
0<\Lambda< \Lambda_{c}\;,
\end{eqnarray}
there are two horizons, $z_{b,c}$; one of them is the black hole horizon, $z_b$, and the other is the cosmic horizon, $z_c$. Now, the singularity is hidden by the black hole horizon, $z_b$.  

$\boldsymbol{\gamma}$. {When}
\begin{eqnarray}
\Lambda=\Lambda_{c}\;,
\end{eqnarray}
the black hole horizon and the cosmic horizon coincide and the singularity is hidden by the overlapping~horizons.

$\boldsymbol{\delta}$. {When}
\begin{eqnarray}
\Lambda>\Lambda_{c}\;,
\end{eqnarray}
both the black hole horizon and the cosmic horizon disappear. Since $g_{00}>0$ and $g_{11}<0$ in this case, the~singularity remains~hidden. 

\subsubsection{$s=\frac{1}{2}$}

$s=\frac{1}{2}$ corresponds to the critical value $\Lambda_{c}$.
\begin{eqnarray}
\Lambda_{c}=\frac{3}{16M^2}\;.
\end{eqnarray}
{We have} the following~results.  

$\boldsymbol{\alpha}$.
 {When}
\begin{eqnarray}
\Lambda\leq 0\;,
\end{eqnarray}
there is only a naked singularity at $z_s=4M$.

$\boldsymbol{\beta}$. {When}
\begin{eqnarray}
0<\Lambda< \Lambda_{c}\;,
\end{eqnarray}
there is no black hole horizon and we are left with a cosmic horizon at $z_c=\sqrt{\frac{3}{\Lambda}}$ and a naked singularity at $z_s=4M$.   

$\boldsymbol{\gamma}$. {When}
\begin{eqnarray}
\Lambda=\Lambda_{c}\;,
\end{eqnarray}
the cosmic horizon and the naked singularity~coincide. 

$\boldsymbol{\delta}$. {When}
\begin{eqnarray}
\Lambda>\Lambda_{c}\;,
\end{eqnarray}
the cosmic horizon disappears while the singularity is hidden in the patch of a de Sitter spacetime. The~patch between the de Sitter horizon and infinity behaves as a one-way membrane, which is just like inside a black~hole.

\subsubsection{$0\leq {s}<\frac{1}{2}$}
In this case, we have the following~conclusions.

$\boldsymbol{\alpha}$. 
{When}
\begin{eqnarray}
\Lambda>0\;,
\end{eqnarray}
there is no black hole horizon and we are left with a cosmic horizon  and a naked~singularity.

$\boldsymbol{\beta}$.
 {When}
\begin{eqnarray}
\Lambda\leq 0\;,
\end{eqnarray}
there is neither a black hole horizon nor a cosmic horizon. We are left with only a naked~singularity.

\subsection{Spacetime Structure of Wormhole in~DAU}
The spacetime for a wormhole in DAU is given by Equation~(\ref{desire}) under the condition that $s>1$. From~the metric, we can calculate the Ricci scalar
\vspace{-6pt}
\begin{adjustwidth}{-\extralength}{0cm}
\begin{eqnarray}
R&=&\frac{2M^2\left(s^2-1\right)}{z^4s^2}\left(1-\frac{2M}{zs}\right)^{s-2}\nonumber\\&&-\frac{2\Lambda}{z^2s^2}\left(1-\frac{2M}{zs}\right)^{-s-1}\left[2z^2s^2-\left(4sM+4s^2M\right)z+4sM^2+M^2+3s^2M^2
\right]\;.
\end{eqnarray}
\end{adjustwidth}

The domain of $z$ is $z\in\left[\frac{2M}{s}, +\infty\right)$. It is apparent that the Ricci scalar is divergent at $z_s=\frac{2M}{s}$. Thus, $z_s=\frac{2M}{s}$ is a physical~singularity. 

Let $\frac{dg_{22}}{dz}=0$; we find that the wormhole throat is located at
\begin{eqnarray}
z_t=\frac{M\left(1+s\right)}{s}\;.
\end{eqnarray}

We note that the throat is determined by the mass and the coupling constant $s$. It has nothing to do with the cosmological constant $\Lambda$. Then, we have the following~conclusions.

$\boldsymbol{\alpha}$.
 {When}
\begin{eqnarray}
\Lambda<0\;,
\end{eqnarray}
we have
\begin{eqnarray}
g_{00}<0\;, \ \ \ \ g_{11}>0\;,
\end{eqnarray}
for arbitrary $z$ within the domain $z\in\left(\frac{2M}{s}, +\infty\right)$.
Therefore, the spacetime corresponds to a traversable wormhole and a naked singularity in the background of anti-de Sitter spacetime. We note that Lu et al~\cite{lu:2024} find that the wormhole solutions in the background of an anti-de Sitter universe can be constructed in the context of the higher derivative~gravity.

$\boldsymbol{\beta}$.
{There are two} critical values, $\Lambda_{c1}$ and $\Lambda_{c2}$.

$\Lambda_{c1}$ is determined by
\begin{eqnarray}
g_{00}=0\;,\ \ \ \ \ \  \frac{d{g_{22}}}{dz}=0\;.
\end{eqnarray}

 Then, we have
\begin{eqnarray}
\Lambda_{c1}=\frac{3s^2\left(s-1\right)^{2s-1}}{M^2\left(s+1\right)^{2s+1}}\;.
\end{eqnarray}

$\Lambda_{c2}$ is determined by
\begin{eqnarray}
g_{00}=0\;,\ \ \ \ \ \  \frac{d{g_{00}}}{dz}=0\;.
\end{eqnarray}

Thus, we have
\begin{eqnarray}
\Lambda_{c2}=\frac{3s^2\left(2s-1\right)^{2s-1}}{M^2\left(2s+1\right)^{2s+1}}\;.
\end{eqnarray}

We have $\Lambda_{c1}<\Lambda_{c2}$ because $s>1$.

In general, there are two horizons in this spacetime for positive $\Lambda$. One of them is the black hole event horizon $z_{b}$ and the other is the cosmic horizon $z_c$. They correspond to the two real and positive roots of the following equation:
\begin{eqnarray}
g_{00}=0\;.
\end{eqnarray}

Thus, in this case, we have both a black hole and a wormhole throat. 
Then, when
\begin{eqnarray}
0<\Lambda< \Lambda_{c1}\;,
\end{eqnarray}
we have
\begin{eqnarray}
z_b<z_t<z_c\;.
\end{eqnarray}

It reveals that the throat is outside of the black hole. A~free-falling observer far away from the black hole first goes through the wormhole throat and then passes through the black hole event horizon, and eventually, he/she disappears at the black hole singularity $z_s$. On~the other hand, if~we split the spacetime into two parts at $z=z_t$ and glue the two identical parts  $z\in\left[z_t, +\infty\right)$ together at $z=z_t$, we arrive at a traversable~wormhole.

$\boldsymbol{\gamma}$. {When}
\begin{eqnarray}
\Lambda=\Lambda_{c1}\;,
\end{eqnarray}
the black hole horizon and the wormhole throat coincide.
Like for the Schwarzschild wormhole, this wormhole is not~traversable.

$\boldsymbol{\delta}$. {When}
\begin{eqnarray}
\Lambda_{c1}<\Lambda<\Lambda_{c2}\;,
\end{eqnarray}
we have
\begin{eqnarray}
z_t<z_b<z_c\;.
\end{eqnarray}

This reveals that the throat is inside of the black~hole.

$\boldsymbol{\epsilon}$. {When}
\begin{eqnarray}
\Lambda=\Lambda_{c2}\;,
\end{eqnarray}
the black hole horizon and the cosmic horizon~coincide.

$\boldsymbol{\zeta}$. {When}
\begin{eqnarray}
\Lambda>\Lambda_{c2}\;,
\end{eqnarray}
we have
\begin{eqnarray}
g_{00}>0\;,\ \ \ \ g_{11}<0\;,
\end{eqnarray}
for arbitrary $r$. Therefore, there are no horizons in this~spacetime. 

\subsection{Spacetime Structure of Exponential Wormhole in~DAU}
The spacetime for a wormhole in DAU is given by Equation~(\ref{desire}) under the condition that $s>1$. From~the metric, we can calculate the Ricci scalar
\vspace{-6pt}
\begin{adjustwidth}{-\extralength}{0cm}
\begin{eqnarray}
R&=&\frac{2M^2\left(s^2-1\right)}{z^4s^2}\left(1-\frac{2M}{zs}\right)^{s-2}\nonumber\\&&-\frac{2\Lambda}{z^2s^2}\left(1-\frac{2M}{zs}\right)^{-s-1}\left[2z^2s^2-\left(4sM+4s^2M\right)z+4sM^2+M^2+3s^2M^2
\right]\;.
\end{eqnarray}
\end{adjustwidth}

The domain of $z$ is $z\in\left[\frac{2M}{s}, +\infty\right)$. It is apparent that the Ricci scalar is divergent at $z_s=\frac{2M}{s}$. Thus, $z_s=\frac{2M}{s}$ is a physical~singularity. 

Let $\frac{dg_{22}}{dz}=0$; we find that the wormhole throat is located at
\begin{eqnarray}
z_t=\frac{M\left(1+s\right)}{s}\;.
\end{eqnarray}

We note that the throat is determined by the mass and the coupling constant $s$. It has nothing to do with the cosmological constant $\Lambda$. Then, we have the following~conclusions.

$\boldsymbol{\alpha}$.
 {When}
\begin{eqnarray}
\Lambda<0\;,
\end{eqnarray}
we have
\begin{eqnarray}
g_{00}<0\;, \ \ \ \ g_{11}>0\;,
\end{eqnarray}
for arbitrary $z$ within the domain $z\in\left(\frac{2M}{s}, +\infty\right)$.
Therefore, the spacetime has a traversable wormhole and a naked singularity in the background of anti-de Sitter~spacetime. 

$\boldsymbol{\beta}$.
 {There are two} critical values, $\Lambda_{c_1}$ and $\Lambda_{c_2}$.
$\Lambda_{c_1}$ is determined by
\begin{eqnarray}
g_{00}=0\;,\ \ \ \ \ \  \frac{d{g_{22}}}{dz}=0\;.
\end{eqnarray}

 Then, we have
\begin{eqnarray}
\Lambda_{c1}=\frac{3s^2\left(s-1\right)^{2s-1}}{M^2\left(s+1\right)^{2s+1}}\;.
\end{eqnarray}

$\Lambda_{c2}$ is determined by
\begin{eqnarray}
g_{00}=0\;,\ \ \ \ \ \  \frac{d{g_{00}}}{dz}=0\;.
\end{eqnarray}

Thus, we have
\begin{eqnarray}
\Lambda_{c2}=\frac{3s^2\left(2s-1\right)^{2s-1}}{M^2\left(2s+1\right)^{2s+1}}\;.
\end{eqnarray}

We have $\Lambda_{c1}<\Lambda_{c2}$ because $s>1$.

In general, there are two horizons in this spacetime for positive $\Lambda$. One of them is the black hole event horizon $z_{b}$ and the other is the cosmic horizon $z_c$. They correspond to the two real and positive roots of the following equation:
\begin{eqnarray}
g_{00}=0\;.
\end{eqnarray}

Thus, in this case, we have both a black hole and a wormhole throat. 
Then, when
\begin{eqnarray}
0<\Lambda< \Lambda_{c1}\;,
\end{eqnarray}
we have
\begin{eqnarray}
z_b<z_t<z_c\;.
\end{eqnarray}

It reveals that the throat is outside of the black hole. A~free-falling observer far away from the black hole first goes through the wormhole throat and then passes through the black hole event horizon; eventually, he/she disappears at the black hole singularity $z_s$. On~the other hand, if~we split the spacetime into two parts at $z=z_t$ and glue the two identical parts  together $z\in\left[z_t, +\infty\right)$ at $z=z_t$, we arrive at a traversable~wormhole.

$\boldsymbol{\gamma}$. {When}
\begin{eqnarray}
\Lambda=\Lambda_{c1}\;,
\end{eqnarray}
the black hole horizon and the wormhole throat coincide. 
Like for the Schwarzschild wormhole, this wormhole is not~traversable.

$\boldsymbol{\delta}$. {When}
\begin{eqnarray}
\Lambda_{c1}<\Lambda<\Lambda_{c2}\;,
\end{eqnarray}

we have
\begin{eqnarray}
z_t<z_b<z_c\;.
\end{eqnarray}

This reveals that the throat is inside of the black~hole.

$\boldsymbol{\epsilon}$. {When}
\begin{eqnarray}
\Lambda=\Lambda_{c2}\;,
\end{eqnarray}
the black hole horizon and the cosmic horizon~coincide.

$\boldsymbol{\zeta}$. {When}
\begin{eqnarray}
\Lambda>\Lambda_{c2}\;,
\end{eqnarray}
we have
\begin{eqnarray}
g_{00}>0\;,\ \ \ \ g_{11}<0\;,
\end{eqnarray}
for arbitrary $r$. Therefore, there are no horizons in this~spacetime. 

\section{The Scalar Potential for the Quintessence~Field }\label{sec7}

Both the JNW naked singularity solution and the JNW wormhole solution  satisfy the Einstein equations with the massless scalar field (with respect to quintessence and phantom, respectively). Embedding them in the background of DAU, the~$\Lambda$  term is present in the metric. We expect that the solutions obey the Einstein equations with a massive scalar field. To~be specific, we require a scalar potential. But~we do not want other unnecessary coupling constants to be present in the potential except for $\Lambda$ and $s$. In~this section, we demonstrate that the JNW naked singularity in DAU, Equation~(\ref{desire}), satisfies the Einstein equations with the massive quintessence field
\begin{equation}
R_{\mu\nu}=-\nabla_{\mu}\phi\nabla_{\nu}\phi-g_{\mu\nu}V\left(\phi\right)\;,
\end{equation}
and the equation of motion for quintessence
\begin{equation}
\nabla^2\phi-V_{,\phi}=0\;.
\end{equation}

Here, $V$ is the quintessence potential to be determined.  Substituting Equation~(\ref{desire}) into the field equations, we obtain the scalar field
\begin{eqnarray}
\phi&=&\frac{\sqrt{2}}{2}\sqrt{1-s^2}\ln\left(1-\frac{2M}{sz}\right)\;,
\end{eqnarray}
and the quintessence potential
\begin{eqnarray}\label{pot-Q}
V&=&\frac{\Lambda}{6}\left[\left(2s^2+3s+1\right)e^{-\frac{\phi\sqrt{2-2s^2}}{1+s}}+\left(4-4s^2\right)e^{\frac{s\phi\sqrt{2}}{\sqrt{1-s^2}}}\right.\nonumber\\&&\left.+\left(2s^2-3s+1\right)e^{-\frac{\phi\sqrt{2-2s^2}}{{s-1}}}\right]\;.
\end{eqnarray}

The potential has the desirable feature that it has only two coupling constants, $\Lambda$ and~$s$. It is interesting to conduct a discussion on the behavior of the potential with respect to the parameter $s$.  In~particular, when $s=0$, we have
\begin{eqnarray}
V&=&\frac{\Lambda}{6}\left(4+e^{-\phi \sqrt{2}}+e^{\phi\sqrt{2}}\right)\;,
\end{eqnarray}
while when $s=1/2$, we have
\begin{eqnarray}
V&=&\frac{\Lambda}{2}\left(e^{-\phi \frac{\sqrt{6}}{3}}+e^{\phi\frac{\sqrt{6}}{3}}\right)\;.
\end{eqnarray}

They are positive defined and symmetric with respect to $\phi$.  In~the more general case,  $0<s<1/2$,  the~potential is always positive defined. On~the contrary, if~ $1/2<s<1$, the~potential goes to negative infinity when $\phi=+\infty$.  This is forbidden by the quantum theory of fields. Finally, when $s=1$, we have $V=\Lambda$, which is just the cosmological constant. To~sum up, the~parameter $s$ is constrained to be
\begin{eqnarray}
0\leq s\leq 1/2\;,
\end{eqnarray}
and
\begin{eqnarray}
s=1\;.
\end{eqnarray}

\begin{figure}[H]
	%\centering
	\includegraphics[scale=0.75]{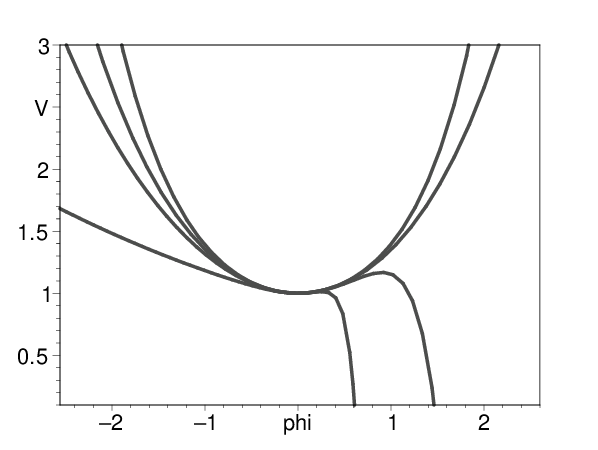}
	\caption{The quintessence potential $V$ for $\Lambda=1$ and $s=0.1,\  0.5,\  0.7,\ 0.95$ from top to bottom, respectively. When $0\leq s\leq 1/2$,  the~potential is always positive defined. On~the contrary, \mbox{if~ $1/2<s<1$, the}~potential goes to negative infinity at $\phi=+\infty$.} 
	\label{quinpot.eps}
\end{figure}

In Figure~\ref{quinpot.eps}, we plot the quintessence potential $V$ for $\Lambda=1$ and $s=0.1,\  0.5,\  0.7,\ 0.95$ from top to bottom, respectively. When $0\leq s\leq 1/2$,  the~potential is always positive defined. In~contrast, if~$1/2<s<1$, the~potential goes to negative infinity at $\phi=+\infty$.

\vspace{-10pt}
\unskip

\section{The Scalar Potential for the Phantom~Field}\label{sec8}
Following the same method in Section 7,  we demonstrate in this section that the wormhole solution in DAU satisfies the Einstein equations with massive phantom field
\begin{equation}\label{EEP}
R_{\mu\nu}=\nabla_{\mu}\phi\nabla_{\nu}\phi-g_{\mu\nu}V\left(\phi\right)\;,
\end{equation}
and
\begin{equation}\label{eomp}
\nabla^2\phi+V_{,\phi}=0\;.
\end{equation}

Here, $V$ is the phantom potential to be determined.  Substituting Equation~(\ref{desire}) into the field equations, we obtain the expression of the phantom  field
\begin{eqnarray}
\phi&=&\frac{\sqrt{2}}{2}\sqrt{s^2-1}\ln\left(1-\frac{b}{z}\right)\;,
\end{eqnarray}
while the scalar potential for the phantom is
\begin{eqnarray}\label{pot-p}
V&=&\frac{\Lambda}{6}\left[\left(2s^2+3s+1\right)e^{\frac{\phi\sqrt{2s^2-2}}{1+s}}+\left(4-4s^2\right)e^{\frac{s\phi\sqrt{2}}{\sqrt{s^2-1}}}\right.\nonumber\\&&\left.+\left(2s^2-3s+1\right)e^{\frac{\phi\sqrt{2s^2-2}}{{s-1}}}\right]\;.
\end{eqnarray}

The potential is always positive for $s>1$. When $\phi=-\infty$,  the~potential is asymptotically vanishing, and when $\phi=+\infty$, it approaches positive infinity. There is a local maximum at $\phi=0$ and a local minimum at
\begin{eqnarray}
\phi&=&\frac{\sqrt{2}}{2}\sqrt{s^2-1}\ln\left(\frac{2s+1}{2s-1}\right)\;.
\end{eqnarray}

\begin{figure}[H]
	%\centering
	\includegraphics[width=8cm,height=6cm]{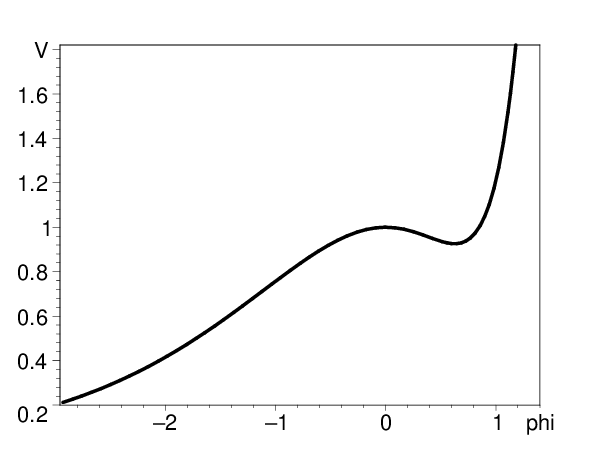}
	\caption{The phantom potential $V$ for $\Lambda=1$ and $s=2$. The~potential is asymptotically vanishing at $\phi=-\infty$ and approaches positive infinity at $\phi=+\infty$. There is a local maximum at $\phi=0$ and a local minimum at $\phi=\frac{\sqrt{6}}{2}\ln\frac{5}{3}$.} 
	\label{pot.eps}
\end{figure}

As an example, we plot the potential with respect to $\phi$ for $\Lambda=1$ and $s=2$ in Figure~\ref{pot.eps}. There is a local maximum at $\phi=0$ and a local minimum at $\phi=\frac{\sqrt{6}}{2}\ln\frac{5}{3}$.

\vspace{-8pt}

In the case of $s=+\infty$, we obtain the metric
\begingroup\makeatletter\def\f@size{9.6}\check@mathfonts
\begin{eqnarray}\label{lambda-1}
dl^2&=&-\left(e^{-\frac{2M}{z}}-\frac{1}{3}\Lambda z^2 e^{\frac{2M}{z}}\right)dT^2+\left(e^{-\frac{2M}{z}}-\frac{1}{3}\Lambda z^2 e^{\frac{2M}{z}}\right)^{-1}dz^2+z^2e^{\frac{2M}{z}}d\Omega^2\;,
\end{eqnarray}
\endgroup
the phantom field
\begin{eqnarray}\label{JNW-PP}
\phi&=&-\frac{\sqrt{2}M}{z}\;,
\end{eqnarray}
and the phantom potential
\begin{eqnarray}\label{phantompot}
V&=&\frac{\Lambda}{3}e^{\sqrt{2}\phi}\left(3-3\sqrt{2}\phi+2\phi^2\right)\;.
\end{eqnarray}

\begin{figure}[H]
	%\centering
	\includegraphics[width=8cm,height=6cm]{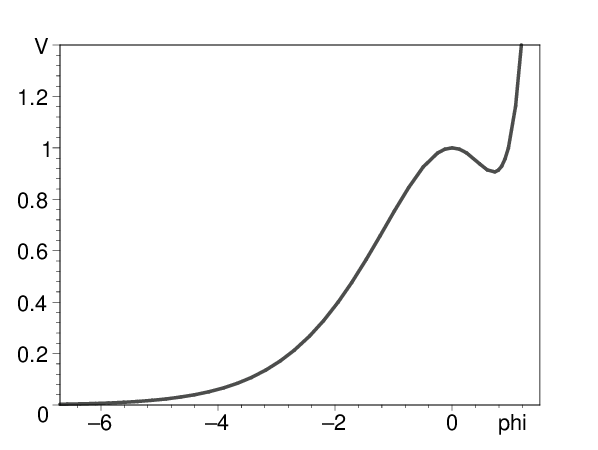}
	\caption{The phantom potential $V$ for $\Lambda=1$ and $s=+\infty$. The~potential is asymptotically vanishing at $\phi=-\infty$ and approaches positive infinity at $\phi=+\infty$. There is a local maximum at $\phi=0$ and a local minimum at $\phi=\frac{\sqrt{2}}{2}$.} 
	\label{pota.eps}
\end{figure}

\textls[-20]{We have checked that the above solution does satisfy the Einstein equations, \mbox{Equation~(\ref{EEP})},  and the phantom field equation, Equation~(\ref{eomp}).  In~Figure~\ref{pota.eps}, we plot the potential with respect to $\phi$ for $\Lambda=1$. The~potential is asymptotically vanishing at $\phi=-\infty$ and approaches positive infinity at $\phi=+\infty$. There is a local maximum at $\phi=0$ and a local minimum at $\phi=\frac{\sqrt{2}}{2}$.}

\vspace{-10pt}
\unskip

\section{Quintessence Potential Is Exactly the Dilaton~Potential}\label{sec9}

Observing the quintessence potential, Equation~(\ref{pot-Q}),  and the phantom potential. \mbox{Equation~(\ref{pot-p})}, we find that they look very similar to the dilaton potential. This is not a coincidence, and~in fact the quintessence potential turns out to be the dilaton potential after redefinition of the coupling constants. To~illustrate this point, we let
\begin{eqnarray}
s=1-\frac{2\alpha^2}{1+\alpha^2}\;,\ \ \ \phi\xrightarrow{}\sqrt{2}\phi\;,\ \ \ \Lambda\xrightarrow{}2\Lambda\;.
\end{eqnarray}

Then, we obtain
\begin{eqnarray}
V&=&\frac{2\Lambda}{3\left(1+\alpha^2\right)^2}\left[\left(3-\alpha^2\right)e^{2\phi\alpha}+8\alpha^2e^{\phi\alpha+\phi/\alpha}\right.\nonumber\\&&\left.+\left(3\alpha^4-\alpha^2\right)e^{-2\phi/\alpha}\right]\;.
\end{eqnarray}

It is exactly the dilaton potential. On~the other hand, with~the replacement of  $\phi$ with $-i\phi$, the~quintessence potential becomes the phantom potential. This motivates us to look for the charged phantom wormholes or charged phantom black holes from the Einstein--Maxwell dilaton black holes by complex~transformations.

\section{Charged Phantom Wormholes and Black~Holes}\label{sec10}
\unskip 
\subsection{JNW Naked Singularity Solution Is~Unphysical}
The charged dilaton black hole solution in the background of DAU is given by~\cite{gao:2004}

\begin{adjustwidth}{-\extralength}{0cm}
%\centering %% If there is a figure in wide page, please release command \centering
\begin{eqnarray}\label{dilaton}
dl^2&=&-\left[\left(1-\frac{r_{+}}{r}\right)\left(1-\frac{r_{-}}{r}\right)^{\frac{1-\alpha^2}{1+\alpha^2}}-\frac{1}{3}\Lambda r^2\left(1-\frac{r_{-}}{r}\right)^{\frac{2\alpha^2}{1+\alpha^2}}\right]dt^2\nonumber\\&&
+\left[\left(1-\frac{r_{+}}{r}\right)\left(1-\frac{r_{-}}{r}\right)^{\frac{1-\alpha^2}{1+\alpha^2}}-\frac{1}{3}\Lambda r^2\left(1-\frac{r_{-}}{r}\right)^{\frac{2\alpha^2}{1+\alpha^2}}\right]^{-1}dr^2+ r^2\left(1-\frac{r_{-}}{r}\right)^{\frac{2\alpha^2}{1+\alpha^2}}d\Omega_2^2\;.
\end{eqnarray}
\end{adjustwidth}
where $r_{+}$, $r_{-}$ are two constants and they are given by~\cite{gib:1988,gar:1991}
\begin{eqnarray}\label{r+-}
r_{+}=m+\sqrt{m^2-\left(1-\alpha^2\right)Q^2}\;,\ \ \ \ r_{-}=\frac{1+\alpha^2}{1-\alpha^2}\left[m-\sqrt{m^2-\left(1-\alpha^2\right)Q^2}\right]\;.
\end{eqnarray}

Here, the constant $\alpha$ governs the strength of the
coupling between the dilaton and the Maxwell field. $m$ and $Q$ stand for the mass and charge of the black hole. $\Lambda$ is understood as the cosmological constant. The~action
leading to the metric is
\begin{eqnarray}
S&=&\int
d^4x\sqrt{-g}\left\{R-2\partial_{\mu}\phi\partial^{\mu}\phi-e^{-2\alpha\phi}F^2
\right.\nonumber\\
&&\left.-\frac{2}{3}\Lambda\frac{1}{\left(1+\alpha^2\right)^2}\left[\alpha^2\left(3\alpha^2-1\right)e^{-2\phi/\alpha}
+\left(3-\alpha^2\right)e^{2\phi\alpha}+8\alpha^2e^{\phi\alpha-\phi/\alpha}\right]\right\}.
\end{eqnarray}

Corresponding to the metric, the~dilaton field and the Maxwell field  are
\begin{equation}
e^{2\alpha\phi}=\left(1-\frac{r_-}{r}\right)^{\frac{2\alpha^2}{1+\alpha^2}}\;,\ \ \  \ F_{01}=\frac{Q}{r^2}\;.
\end{equation}

Comparing the dilaton black hole solution in DAU, Equation~(\ref{dilaton}),  with~the JNW solution in DAU, Equation~(\ref{desire}), and~we perform the following transformations:
\begin{equation}\label{insist}
Q=0\;,\ \ \ \ m=-\bar{m}\;,\ \ \ \ \ \alpha^2=\frac{1-s}{1+s}\;, 
\end{equation}
in Equation~(\ref{r+-}) under the condition that $\bar{m}$ is positive; then, Equation~(\ref{dilaton}) becomes
\begin{eqnarray}
dl^2&=&-\left[\left(1+\frac{2\bar{m}}{sr}\right)^{s}-\frac{1}{3}\Lambda r^2\left(1+\frac{2\bar{m}}{sr}\right)^{1-s}\right]dt^2\nonumber\\&&
+\left[\left(1+\frac{2\bar{m}}{sr}\right)^{s}-\frac{1}{3}\Lambda r^2\left(1+\frac{2\bar{m}}{sr}\right)^{1-s}\right]^{-1}dr^2+ r^2\left(1+\frac{2\bar{m}}{sr}\right)^{1-s}d\Omega_2^2\;.
\end{eqnarray}

Let
\begin{equation}
\bar{m}=-M\;, 
\end{equation}
and the above metric turns out to be the JNW solution in DAU. Now, we realize that the constant $s$ is actually related to the dilaton coupling constant $\alpha$ and the parameter $M$ is the negative of mass for the dilaton black hole. In~brief, \emph{{the JNW solution is actually the neutral dilaton black hole solution but with a negative mass}%MDPI: Please confirm if the italics is unnecessary and can be removed. The following highlights are the same.
}. A~negative mass is not physically allowed. Therefore, we are unable to generate the JNW solution from the dilaton one. If~we insist on performing the above transformations in Equation~(\ref{insist}), the~JNW solution would not be~physical. 

\subsection{Two Double-Horizon Spacetimes Connected by a Timelike~Wormhole}

\textls[-10]{In order to obtain phantom wormholes from the above dilaton black hole solution\mbox{, we let}}
\begin{equation}
\phi\rightarrow{e^{i\frac{\pi}{2}}\phi}\;,\ \ \ \ \alpha\rightarrow{e^{i\frac{\pi}{2}}\alpha}\;, 
\end{equation}
with $i$ the imaginary unit. With~these substitutions, the~action and the metric become
\begin{adjustwidth}{-\extralength}{0cm}
\begin{eqnarray}\label{phantomaction}
S&=&\int
d^4x\sqrt{-g}\left\{R+2\partial_{\mu}\phi\partial^{\mu}\phi-e^{2\alpha\phi}F^2
\right.\nonumber\\
&&\left.-\frac{2}{3}\Lambda\frac{1}{\left(1-\alpha^2\right)^2}\left[\alpha^2\left(3\alpha^2+1\right)e^{-2\phi/\alpha}
+\left(3+\alpha^2\right)e^{-2\phi\alpha}-8\alpha^2e^{-\phi\alpha-\phi/\alpha}\right]\right\}\;,
\end{eqnarray}
\end{adjustwidth}
and 

\begin{adjustwidth}{-\extralength}{0cm}
%\centering %% If there is a figure in wide page, please release command \centering
\begin{eqnarray}
dl^2&=&-\left[\left(1-\frac{r_{+}}{r}\right)\left(1-\frac{r_{-}}{r}\right)^{\frac{1+\alpha^2}{1-\alpha^2}}-\frac{1}{3}\Lambda r^2\left(1-\frac{r_{-}}{r}\right)^{\frac{-2\alpha^2}{1-\alpha^2}}\right]dt^2\nonumber\\&&
+\left[\left(1-\frac{r_{+}}{r}\right)\left(1-\frac{r_{-}}{r}\right)^{\frac{1+\alpha^2}{1-\alpha^2}}-\frac{1}{3}\Lambda r^2\left(1-\frac{r_{-}}{r}\right)^{\frac{-2\alpha^2}{1-\alpha^2}}\right]^{-1}dr^2+ r^2\left(1-\frac{r_{-}}{r}\right)^{\frac{-2\alpha^2}{1-\alpha^2}}d\Omega_2^2\;,
\end{eqnarray}
\end{adjustwidth}
with
\begin{eqnarray}
r_{+}=m+\sqrt{m^2-\left(1+\alpha^2\right)Q^2}\;,\ \ \ \ r_{-}=\frac{1-\alpha^2}{1+\alpha^2}\left[m-\sqrt{m^2-\left(1+\alpha^2\right)Q^2}\right]\;.
\end{eqnarray}
{Now, the }identity of the dilaton field changes from dilaton to phantom:
\begin{equation}
e^{-2\alpha\phi}=\left(1-\frac{r_-}{r}\right)^{\frac{-2\alpha^2}{1-\alpha^2}}\;,
\end{equation}
while the expression of the Maxwell field remains~unchanged.

This is exactly a charged wormhole in the background of DAU. We note that the charged Einstein--Maxwell phantom wormholes in the absence of potential were addressed by Nozawa~\cite{noz:2021a}. The~throat of the wormhole is located at
\begin{eqnarray}
r_T=\frac{{r}_{-}}{1-\alpha^2}=\frac{1}{1+\alpha^2}\left[m-\sqrt{m^2-\left(1+\alpha^2\right)Q^2}\right]\;.
\end{eqnarray}

It is then required that the coupling constant satisfy
\begin{eqnarray}
\alpha^2\leq{1}\;,
\end{eqnarray}
in order that $r_{-}$ be physical. We see that
\begin{eqnarray}
r_{T}<r_{-}<r_{+}\;.
\end{eqnarray}

Therefore, the~throat is always inside the inner horizon and it is timelike such that the wormhole is traversable. Observations tell us that the cosmological constant is extremely small and the resulting cosmic horizon is of the order of the Hubble scale. So in general, we can safely neglect the effect of the cosmological constant for local physics. This means that we can neglect the $\Lambda$ term in the metric. Then, we are left with two horizons; one of the them is the black hole event horizon $r_{+}$ and the other is the black hole inner horizon $r_{-}$. The~throat is located inside the inner horizon. By~splitting the spacetime into two parts at $r_T$ and gluing the two identical parts of $r\in\left[r_T, +\infty\right)$ at $r=r_T$, we arrive at a traversable wormhole. A~free-falling observer approaching the black hole first passes through the outer event horizon, then through the inner horizon, wormhole throat, replicated inner horizon, replicated outer horizon, and~finally, he/she reaches another replicated universe. Now, two double-horizon spacetimes are connected by a timelike wormhole.     

\subsection{Two Black--White Hole Spacetimes Connected by a Spacelike~Wormhole}  
When $\alpha=0$,  the~solution reduces to the standard Reissner--Nordstr${\ddot{\rm{o}}}$m--de Sitter solution of Einstein--Maxwell theory. However, when
\begin{eqnarray}
\alpha=1\;,\ \ \Longleftrightarrow{} \ \ \ s=+\infty\;,
\end{eqnarray}
we have
\begin{eqnarray}\label{sinfinity}
dl^2&=&-\left[\left(1-\frac{\bar{r}_{+}}{r}\right)e^{-\frac{\bar{r}_-}{r}}-\frac{1}{3}\Lambda r^2e^{\frac{\bar{r}_-}{r}}\right]dt^2\nonumber\\&&
+\left[\left(1-\frac{\bar{r}_{+}}{r}\right)e^{-\frac{\bar{r}_-}{r}}-\frac{1}{3}\Lambda r^2e^{\frac{\bar{r}_-}{r}}\right]^{-1}dr^2+ r^2e^{\frac{\bar{r}_-}{r}}d\Omega_2^2\;,
\end{eqnarray}
with
\begin{eqnarray}
\bar{r}_{+}=m+\sqrt{m^2-2Q^2}\;,\ \ \ \ \bar{r}_{-}=m-\sqrt{m^2-2Q^2}\;,
\end{eqnarray}
and the phantom field,
\begin{eqnarray}
\phi=-\frac{\bar{r}_{-}}{2r}\;.
\end{eqnarray}

The expression of the Maxwell field is unchanged. The~corresponding potential is
\begin{eqnarray}
V=\frac{2}{3}\Lambda\left(3+6\phi+4\phi^2\right)e^{-2\phi}\;,
\end{eqnarray}
by letting $\alpha\rightarrow{1}$ in the phantom potential of action Equation~(\ref{phantomaction}). 
It is exactly the phantom potential given by Equation~(\ref{phantompot}) after the following transformations:
\begin{eqnarray}
\phi\xrightarrow{}-\phi/\sqrt{2}\;,\ \ \ \Lambda\xrightarrow{}\frac{1}{2}\Lambda\;.
\end{eqnarray}

As mentioned earlier, the~cosmological constant is extremely small. So we neglect the $\Lambda$ term in the metric. Then, we are left with only one black hole event horizon $\bar{r}_{+}$ in spacetime. There is a throat located at
\begin{eqnarray}
r_T=\frac{\bar{r}_{-}}{2}\;.
\end{eqnarray}

Because
\begin{eqnarray}
r_{T}<\bar{r}_{+}\;,
\end{eqnarray}
we see that the throat is always inside the black hole and it is spacelike.  We realize that this is an example that Simpson and Visser~\cite{sim:2019} and Nojiri, Odintsov, and Folomeev~\cite{nojiri:2024} discussed very recently. In~order to show this point, we make the coordinate transformation \mbox{$r\rightarrow{x}$} as follows:
\begin{eqnarray}
r^2e^{\frac{\bar{r}_{-}}{r}}=\frac{e^2}{4}\left(x^2+\bar{r}_{-}^2\right)\;,
\end{eqnarray}
such that when $x=0$, we have $r=\frac{\bar{r}_{-}}{2}$, and when $x=\pm\infty$, we have $r=\infty$. Namely, $x=0$ corresponds to the wormhole throat. The~domain of $x$ is $x\in (-\infty\;,+\infty)$.
Then, we obtain
\begin{eqnarray}
dl^2&=&-\left(1-\frac{\bar{r}_{+}}{r}\right)e^{-\frac{\bar{r}_-}{r}}dt^2\nonumber\\&&
+\left(1-\frac{\bar{r}_{+}}{r}\right)^{-1}\left(1-\frac{\bar{r}_{-}}{2r}\right)^{-2}\left(1+\frac{\bar{r}_{-}^2}{x^2}\right)^{-1}dx^2+\left(x^2+\bar{r}_{-}^2\right)d\Omega_2^2\;,
\end{eqnarray}
after re-scaling the time $t$ and the line element $l$ as follows:
\begin{eqnarray}
t\rightarrow \frac{1}{2}et\;,\ \ \ \ l\rightarrow \frac{1}{2}el\;,
\end{eqnarray}

There are two event horizons in this spacetime, $x=\pm\sqrt{4\bar{r}_{+}^2 e^{\frac{\bar{r}_{-}}{\bar{r}_{+}}-2}-\bar{r}_{-}^2}$. 
One of them is a white hole, instead of a black hole in another universe. An~observer falling into the black hole in our universe first goes through the black hole event horizon, then the wormhole, and~then the white hole event horizon, eventually arriving at another universe. One
can consider the inverse process, that is, an~observer falling into the black hole in another universe would eventually appear in
our universe from the white hole. The~Penrose diagram for the spacetimes of a black--white hole connected by a wormhole can be found in Ref.~\cite{sim:2019}.

\section{Conclusions and~Discussion}\label{sec11}  
Starting from the well-known JNW naked singularity solution, we obtain a new traversable wormhole provided that $s>2$ by using a complex transformation method. The~method is so interesting that we can expect some other new  phantom black holes or phantom wormholes to be derived from the known quintessence black holes. On~the other hand, when $s\rightarrow{+\infty}$, we obtain the well-known exponential metric. To~our knowledge, no one has ever found the corresponding energy momentum tensor. Now, we find that it can be contributed by a massless phantom field. Then, we embed both the JNW naked singularity and the wormhole solution in the background of DAU in an isotropic coordinate system and in a static coordinate system, respectively. The~resulting spacetime structure is very rich. For~example, there exist both black holes and wormholes in the same spacetime.  We construct the quintessence potential and the phantom potential, respectively, for~the solutions.  It is found that the phantom potential is obtained if we perform a complex transformation on the quintessence potential. By~observing the quintessence potential and the dilaton potential, we find that they are very much similar to each other.  On closer inspection, the~quintessence potential proves to be the dilaton potential. This arouses our pursuit of the JNW solution from the charged dilaton metric. We find that in order to obtain the JNW naked singularity, we must assume that the mass of the dilaton black hole is negative. This is of course forbidden, just as one cannot arbitrarily assume the mass in the Schwarzschild metric to be negative. Then, it tells us that the parameter $b$ should be negative just like the parameter $M$ in the Schwarzschild metric is positive. {Based on the above reasons, we think that the JNW naked singularity solution is unphysical}. {However, it is important to take this point with a grain of salt. In~fact, Mann has demonstrated that, under~certain circumstances, regions of negative energy density can undergo gravitational collapse.  Then, the resultant black hole spacetimes have a negative mass but non-trivial topology~\cite{mann:1997}. On~the other hand, Hull and Mann show that a negative mass black hole in de Sitter universe is allowed in the context of Lovelock gravity~\cite{hull:2023}}.

Carrying out the complex transformation on the charged dilaton metric in DAU, we obtain charged two-horizon spacetimes and black--white hole spacetimes, both being connected by a wormhole. The~spacetime structure of two-horizon spacetimes in DAU is very interesting. It can have four important components, the~cosmic horizon, the~black hole event horizon, the~black hole inner horizon, and the wormhole throat. It is found that the wormhole throat is always inside the black hole inner horizon. Therefore, the~throat is timelike and traversable.  As~for the black--white hole spacetimes, we find that the throat is always inside of the black hole event horizon. By~cutting the spacetime into two parts along the throat and gluing the domains of   $x\in (-\infty\;,0)$ and $x\in (0\;,+\infty)$ together at $x=0$, we arrive at a regular spacetime with two horizons, one of which is the black hole horizon, and the other is the white hole~horizon. 

\clearpage 
{\bf{{Note Added} %MDPI: please check if it should be formatted as note, if so, please add footnote in main part where this note belongs to. And list the note content after back matter.
}}

When the paper appeared on arXiv, we were told by Masato Nozawa that the solution in Equation~\eqref{eq47} has already been derived in Ref.~\cite{noz:2021} (Equation~({31a}%MDPI: Please confirm if this equation refer to this manuscript.
)) by the same~method. 

%%%%%%%%%%%%%%%%%%%%%%%%%%%%%%%%%%%%%%%%%%
\vspace{6pt} 

%%%%%%%%%%%%%%%%%%%%%%%%%%%%%%%%%%%%%%%%%%
\authorcontributions{Conceptualization, C.G.; methodology, C.G. and~J.Q; writing—original draft preparation, C.G.; writing---review and editing, C.G. and J.Q. All authors have read and agreed to the published version of the~manuscript.}

\funding{\textls[10]{This work was supported in part by the National Key R$\&$D Program of China grants \mbox{No. 2022YFF0503404} and No.~2022SKA0110100.}}

\dataavailability{Files used to generate the various graphs presented in this paper are available on~request.}

\acknowledgments{We are very grateful to the referees for the expert
suggestions and the careful corrections,  which have significantly improved the {paper.}}
%The work is supported by the National Key  R$\&$D Program of China grants No. 2022YFF0503404 and No.~2022SKA0110100.}%MDPI: We removed the duplicated information here with the funding section. Please confirm.

\conflictsofinterest{The authors declare no conflicts of interest.}
\begin{adjustwidth}{-\extralength}{0cm}

\reftitle{References}

\PublishersNote{}
\end{adjustwidth}

\begin{thebibliography}{999}
\bibitem{jnw:1968} Janis, A.I.; Newman, E.T.; Winicour, J. Reality of the Schwarzschild Singularity. \emph{{Phys. Rev. Lett.} %MDPI: Newly added information, please confirm the following highlights are the same.
} {\textbf{1968}}, \emph{20}, {878}.

\bibitem{fisher:1948} Fisher, I.Z. Scalar metastatic field with regard for gravitational effects. \emph{Zh. Eksp. Teor. Fiz.} \textbf{1948}, {\emph{18}}, 636–640.

\bibitem{wyman:1981} Wyman, M. Static Spherically Symmetric Scalar Fields in General Relativity. \emph{{Phys. Rev. D}} {\textbf{1981}}, \emph{24}, {839}.

\bibitem{virb:1997} Virbhadra, K.S. Janis-Newman-Winicour and Wyman solutions are the same. \emph{Int. J. Mod. Phys. A} \textbf{1997}, {\emph{12}}, 4831–4836.

\bibitem{agn:1985} Agnese, A.G.; La Camera, M. Gravitation without black holes. {{Phys. Rev. D}} {\textbf{1985}}, \emph{31}, {1280}.

\bibitem{rob:1993} Roberts, M.D. Massless scalar static spheres. \emph{Astrophys. Space Sci.} \textbf{1993}, {\emph{200}}, 331.

\bibitem{bron:1979} Bronnikov, K.A.; Khodunov, A.V. Scalar field and gravitational instability. \emph{Gen. Relativ. Gravit.} \textbf{1979}, {\emph{11}}, 13–18.

\bibitem{chew:2024} Chew, X.Y.; Lim, K.G. Gravitating Scalarons with Inverted Higgs Potential. \emph{Universe} \textbf{2024}, {\emph{10}}, 212.

\bibitem{gyu:2008} Gyulchev, G.N.; Yazadjiev, S.S. Gravitational Lensing by Rotating Naked Singularities. \emph{Phys. Rev. D }\textbf{2008}, {\emph{78}}, 083004.

\bibitem{virb:2002} Virbhadra, K.S.; Ellis, G.F.R. Gravitational lensing by naked singularities. \emph{Phys. Rev. D} \textbf{2002}, {\emph{65}}, 103004.

\bibitem{virb:1998} Virbhadra, K.S.; Narasimha, D.; Chitre, S.M. Role of the scalar field in gravitational lensing. \emph{Astron. Astrophys.} \textbf{1998}, {\emph{337}}, 1–8.

\bibitem{virb:2008} Virbhadra, K.S.; Keeton, C.R. Time delay and magnification centroid due to gravitational lensing by black holes and naked singularities. \emph{Phys. Rev. D} \textbf{2008}, {\emph{77}}, 124014.

\bibitem{gyu:2019} Gyulchev, G.; Nedkova, P.; Vetsov, T.; Yazadjiev, S. Image of the Janis-Newman-Winicour naked singularity with a thin accretion disk. \emph{Phys. Rev. D} \textbf{2019}, {\emph{100}}, 024055.

\bibitem{sub:2020} Sau, S.; Banerjee, I.; SenGupta, S. Imprints of the Janis-Newman-Winicour spacetime on observations related to shadow and accretion. \emph{Phys. Rev. D} \textbf{2020}, {\emph{102}}, 064027.

\bibitem{yang:2015} Yang, L.; Li, Z. Shadow of a dressed black hole and determination of spin and viewing angle. \emph{Int. J. Mod. Phys. D} \mbox{\textbf{2015}, {\emph{25}}, 1650026.}

\bibitem{tak:2004} Takahashi, R. Shapes and positions of black hole shadows in accretion disks and spin parameters of black holes. \emph{J. Korean Phys. Soc. }\textbf{2004}, {\emph{45}}, S1808--S1812.

\bibitem{chow:2012} Chowdhury, A.N.; Patil, M.; Malafarina, D.; Joshi, P.S. Circular geodesics and accretion disks in Janis-Newman-Winicour and Gamma metric. \emph{Phys. Rev. D} \textbf{2012}, {\emph{85}}, 104031.

\bibitem{pal:2023} Pal, K.; Pal, K.; Shaikh, R.; Sarkar, T. A rotating modified JNW spacetime as a Kerr black hole mimicker. \emph{J. Cosmol. Astropart. Phys.} \textbf{2023}, {\emph{11}}, 060.

\bibitem{sta:1} Zhdanov, V.; Stashko, O. Static spherically symmetric configurations with N non-linear scalar fields: Global and asymptotic properties. \emph{Phys. Rev. D} \textbf{2020}, \emph{101}, {064064}.

\bibitem{sta:2} Stashko, O.S.; Zhdanov, V.I.; Alexandrov, A.N. Thin accretion discs around spherically symmetric configurations with nonlinear scalar fields. \emph{Phys. Rev. D} \textbf{2021}, \emph{104}, {104055}.

\bibitem{sta:3} Stashko, O.S.; Savchuk, O.V.; Zhdanov, V.I. Quasi-normal modes of naked singularities in presence of non-linear scalar fields. \emph{Phys. Rev. D} {\textbf{2024}}, \emph{109}, {024012}.

\bibitem{matos:1}
Matos, T.; Nuñez, D. Rotating Scalar Field Wormhole. 
\emph{Class. Quant. Gravit.} \textbf{2006}, \emph{23}, 4485--4495.

\bibitem{matos:2}
Matos, T. Class of Einstein-Maxwell phantom Fields: Rotating and Magnetized Wormholes. 
\emph{Gen. Relativ. Gravit.} \textbf{2010}, \emph{42}, 1969.

\bibitem{matos:3}
Matos, T.; Miranda, G.; Montelongo, N. 
Kerr-like Scalar Field Wormhole.
\emph{Gen. Relativ. Gravit.} \textbf{2014}, \emph{46}, 1613.

\bibitem{matos:4}
Matos, T.; Arturo Ureña-López, L.; Miranda, G. Wormhole Cosmic Censorship.
\emph{Gen. Relativ. Gravit.} \textbf{2016}, \emph{48}, 61.

\bibitem{matos:5}
del Aguila, J.C.; Matos, T. Wormhole Cosmic Censorship: An Analytical Proof.
\emph{Gen. Relativ. Gravit.} \textbf{2019}, \emph{36}, 015018.

\bibitem{matos:6}
Miranda, G.; del Aguila, J.C.; Matos, T. Exact Rotating Magnetic Traversable Wormholes satisfying the Energy Conditions.
\emph{Phys. Rev. D }\textbf{2019}, \emph{99}, 124045.

\bibitem{matos:7}
del Aguila, J.C.; Matos, T. Gravitational perturbations in the Newman-Penrose formalism: Applications to wormholes.
\emph{Phys. Rev. D} \textbf{2021}, \emph{103}, 084033.

\bibitem{matos:8}
del Aguila, J.C.; Matos, T. On the geodesic completeness of a ring wormhole.
\emph{Phys. Rev. D} \textbf{2023}, \emph{107}, 064047.

\bibitem{sad:2013} Sadhu, A.; Suneeta, V. A naked singularity stable under scalar field perturbations. \emph{Int. J. Mod. Phys. D} \textbf{2013}, {\emph{22}}, 1350015.

\bibitem{gib:1988} Gibbons, G.; Maeda, K. Black holes and membranes in higher-dimensional theories with dilaton fields. \emph{Nucl. Phys. B} \textbf{1988}, \mbox{{\emph{298}}, 741}.

\bibitem{gar:1991} Garfinkle, D.; Horowitz, G.; Strominger, A. Charged black holes in string theory. \emph{Phys. Rev.
D }\textbf{1991}, {\emph{43}}, 3140.

\bibitem{caldwell:2002}Caldwell, R.R. A Phantom Menace? Cosmological consequences of a dark energy component with super-negative equation of state. \emph{Phys. Lett. B} \textbf{2002}, {\emph{545}}, 23--29.

\bibitem{pap:1954} Papapetrou, A. Eine Theorie des Gravitationsfeldes mit einer Feldfunktion. \emph{Z. Fur Phys. Bd.} \textbf{1954}, {\emph{139}}, 518.

\bibitem{gao:2004}Gao, C.; Zhang, S.N. Dilaton black holes in the de Sitter or anti---de Sitter universe. \emph{{Phys. Rev. D.}} \textbf{2004}, \emph{70}, {124019}.

\bibitem{noz:2023} Nozawa, M.; Torii, T. Wormhole C metric. \emph{Phys. Rev.
D }\textbf{2023}, \emph{108}, {064036}.

\bibitem{boo:2018}Boonserm, P.; Ngampitipan, T.; Simpson, A.; Visser, M. Exponential metric represents a traversable wormhole. \emph{Phys. Rev. D }\mbox{\textbf{2018}, \emph{98}, {084048}.}

\bibitem{mor:1988}Morris, M.S.; Thorne, K.S. Wormholes in space-time and their use for interstellar travel: A tool for teaching general relativity. \emph{Am. J. Phys. }\textbf{1988}, {\emph{56}}, 395.

\bibitem{mor:1988b} Morris, M.S.; Thorne, K.S.; Yurtsever, U. Wormholes, Time Machines, and the Weak Energy Condition. \emph{Phys. Rev. Lett. }\textbf{1988}, \mbox{{\emph{61}}, 1446. }

\bibitem{vis:1989} Visser, M. Traversable wormholes: Some simple examples. \emph{Phys. Rev. D } \textbf{1989}, {\emph{39}}, 3182.

\bibitem{vis:1989b} Visser, M. Traversable wormholes from surgically modified Schwarzschild space-times. \emph{Nucl. Phys. B} \textbf{1989}, {\emph{328}}, 203.

\bibitem{tak:2023}Tangphati, T.; Muniz, C.R.; Pradhan, A.; Banerjee, A. Traversable wormholes in Rastall-Rainbow gravity. \emph{Phys. Dark Univ. }\textbf{2023}, {\emph{42}}, 101364.

\bibitem{kob:2012} Kobayashi, T.; Motohashi, H.; Suyama, T. Black hole perturbation in the most general scalar-tensor theory with second-order field equations. II. The even-parity sector. \emph{Phys. Rev. D} \textbf{2012}, {\emph{85}}, 084025.

\bibitem{gao:2022}Gao, C.; Qiu, J. On black holes with scalar hairs. \emph{Gen. Relativ. Gravit. } \textbf{2022}, {\emph{54}}, 158.

\bibitem{yil:1958} Yilmaz, H. New approach to general relativity. \emph{Phys. Rev. }\textbf{1958}, {\emph{111}}, 1417–1426.

\bibitem{yil:1971} Yilmaz, H. New theory of gravitation. \emph{Phys. Rev. Lett.} \textbf{1971}, \emph{27}, {1399}.

\bibitem{yil:1973} Yilmaz, H. New approach to relativity and gravitation. \emph{Ann. Phys. } \textbf{1973}, {\emph{81}}, 179–200.

\bibitem{cla:1973} Roger E. Clapp, Preliminary quasar model based on the Yilmaz exponential metric. \emph{Phys. Rev. D} \textbf{1973}, {\emph{7}}, 345–355.

\bibitem{1}Rastall, P. Gravity without geometry. \emph{Am. J. Phys.} \textbf{1975}, {\emph{43}}, 591–595.

\bibitem{2} Fennelly, A.J.; Pavelle, R. Nonviability of Yilmaz’ gravitation theories and his
criticisms of Rosen’s gravitation theory. Print-76-0905. %MDPI: Please provide more information about the article type, such as book (please provide the name and location of the publisher); online resource (please provide the URL of the website and the date it was accessed (Date Month Year)); or journal article (please provide the name of the journal, the year and volume in which it was published, and the page number). Please refer to https://www.mdpi.com/authors/references for full reference formatting guides.

\bibitem{3} Misner, C.W. Yilmaz cancels Newton. \emph{Nuovo Cim. B }\textbf{1999}, {\emph{114}}, 1079{.} %MDPI: We removed extra information, please confirm.

\bibitem{4} Alley, C.O.; Aschan, P.K.; Yilmaz, H. Refutation of C.W. Misner’s claims in his
article ‘Yilmaz cancels Newton’, gr-qc/9506082.
\bibitem{5} Robertson, S.L. X-Ray novae, event horizons, and the exponential metric. \emph{ Astrophys. J.}
\textbf{1999}, \emph{515}, {365}.

\bibitem{6} Robertson, S.L. Bigger bursts from merging neutron stars. \emph{ Astrophys. J.}
{\textbf{1999}}, \emph{517}, {L117}.

\bibitem{7} Ibison, M. The Yilmaz cosmology. \emph{AIP Conf. Proc. }\textbf{2006}, {\emph{822}}, 181.

\bibitem{8} Ibison, M. Cosmological test of the Yilmaz theory of gravity.
\emph{Class. Quant. Gravit.} \textbf{2006}, {\emph{23}}, 577.

\bibitem{9} Ben-Amots, N.
Relativistic exponential gravitation and exponential potential of electric charge.
\emph{Found. Phys.} \textbf{2007}, {\emph{37}}, 773.

\bibitem{10} Svidzinsky, A.A.
Vector theory of gravity in Minkowski space-time: Flat universe without black holes. \emph{arXiv} \textbf{2009},
arXiv:0904.3155.

\bibitem{11} Martinis, M.; Perkovic, N.
Is exponential metric a natural space-time metric of Newtonian gravity? \emph{arXiv }\textbf{2010}, 
arXiv:1009.6017.

\bibitem{12} Ben-Amots, N. Some features and implications of exponential gravitation.
\emph{J. Phys. Conf. Ser. }\textbf{2011}, {\emph{330}}, 012017.

\bibitem{13} Svidzinsky, A.A. Vector theory of gravity: Universe without black holes and solution of dark energy problem.
\emph{Phys. Scr.} \textbf{2017}, \mbox{{\emph{92}}, 125001.}

\bibitem{14} Aldama, M.E. The gravity apple tree. \emph{J. Phys. Conf. Ser. }\textbf{2015}, {\emph{600}}, 012050.

\bibitem{15}Robertson, S.L. MECO in an exponential metric. \emph{arXiv} \textbf{2016}, arXiv:1606.01417.

\bibitem{sim:2021}Simpson, A. Traversable Wormholes, Regular Black Holes, and Black-Bounces. \emph{arXiv }\textbf{2021}, arXiv:2104.14055.

\bibitem{1a} Lobo, F.S.N.; Crawford, P. Linearized stability
analysis of thin shell wormholes with a cosmological
constant. \emph{Class. Quant. Gravit.} \textbf{2004}, {\emph{21}}, 391--404.

\bibitem{2a} Lemos, J.P.S.; Lobo, F.S.N. Plane symmetric thin-shell wormholes: Solutions and stability. \emph{Phys. Rev. D} \textbf{2008}, \emph{78}, 044030.

\bibitem{3a}Li, A.C.; Xu, W.L.; Zeng, D.F. Linear Stability Analysis of Evolving Thin Shell Wormholes. \emph{J. Cosmol. Astropart. Phys.} \textbf{2019}, \mbox{\emph{1903}, 016.}

\bibitem{4a} Hochberg, D.; Kephart, T.W. Wormhole cosmology and the horizon problem. \emph{Phys. Rev. Lett.}
\textbf{1993}, {\emph{70}}, 2665--2668.

\bibitem{5a}Kim, S.W. Evolution of Cosmological Horizons of
Wormhole Cosmology. \emph{Int. J. Mod. Phys. D} \textbf{2020}, {\emph{29}}
, 2050079.

\bibitem{6a}Bhawal, B.; Kar, S. Lorentzian wormholes in
Einstein-Gauss-Bonnet theory. \emph{Phys. Rev. D} \textbf{1992}, {\emph{46}},
2464--2468.

\bibitem{7a} Hochberg, D. Lorentzian wormholes in higher order
gravity theories. \emph{Phys. Lett. B }\textbf{1990}, {\emph{251}}, 349--354.

\bibitem{8a} Agnese, A.G.; La Camera, M. Wormholes in the
Brans-Dicke theory of gravitation. \emph{Phys. Rev. D} \textbf{1995}, {\emph{51}}, 2011--2013.

\bibitem{9a}Jusufi, K.; Banerjee, A.; Ghosh, S.G. Wormholes
in 4D Einstein–Gauss–Bonnet gravity. \emph{Eur. Phys. J. C}
\textbf{2020}, {\emph{80}}, 698.

\bibitem{10a}Huang, H.; Lu, H.; Yang, J. Bronnikov-like Wormholes in Einstein-Scalar Gravity. \emph{arXiv }\textbf{2010}, arXiv:2010.00197.

\bibitem{11a}Ibadov, R.; Kleihaus, B.; Kunz, J.; Murodov, S. Wormholes in Einstein-scalar-Gauss-Bonnet
theories with a scalar self-interaction potential. \emph{Phys. Rev. D }\textbf{2020}, {\emph{102}}, 064010.

\bibitem{12a}Kanti, P.; Kleihaus, B.; Kunz, J. Wormholes in Dilatonic Einstein-Gauss-Bonnet Theory. \emph{Phys. Rev. Lett.} \textbf{2011}, {\emph{107}}, 271101.

\bibitem{13a}Rosa, J.L.; Lemos, J.P.S.; Lobo, F.S.N. Wormholes
in generalized hybrid metric-Palatini gravity obeying the
matter null energy condition everywhere. \emph{Phys. Rev. D} \textbf{2018}, {\emph{98}}, 064054.

\bibitem{14a}Cariglia, M.; Gibbons, G.W. Levy-Leblond
fermions on the wormhole. \emph{arXiv} \textbf{2018}, arXiv:1806.05047.

\bibitem{15a}Blazquez-Salcedo, J.L.; Knoll, C.; Radu, E.
Traversable wormholes in Einstein-Dirac-Maxwell theory. \emph{Phys. Rev. Lett.} \textbf{2021}, \mbox{{\emph{126}}, 101102.}

\bibitem{16a}Sahoo, P.K.; Moraes, P.H.R.S.; Sahoo, P.; Ribeiro, G. Phantom fluid supporting traversable wormholes in alternative gravity with extra material terms. \emph{Int. J. Mod. Phys. D} \textbf{2018}, {\emph{27}}, 1950004.

\bibitem{17a}Parsaeia, F.; Rastgoob, S. Wormhole solutions with a polynomial equation-of-state and minimal violation of the null energy condition.
\emph{Eur. Phys. J. C} \textbf{2020}, {\emph{80}}, 366.

\bibitem{18a}Lobo, F.S.N.; Rodrigues, M.E.; Silva, M.V.D.S.; Simpson, A.; Visser, M. Novel black-bounce spacetimes: Wormholes, regularity, energy conditions, and causal structure. \emph{Phys. Rev. D} \textbf{2021}, 
{\emph{103}}, 084052.

\bibitem{19a}Di Grezia, E.; Battista, E.; Manfredonia, M.; Miele, G. Spin, torsion and violation of null energy condition in traversable wormholes. \emph{Eur. Phys. J. Plus} \textbf{2017}, {\emph{132}}, 537.

\bibitem{lu:2024}Lu, M.; Yang, J.; Mann, R.B. Gravitational Wormholes. \emph{Universe }\textbf{2024}, {\emph{10}}, 257.

\bibitem{noz:2021a} Nozawas, M. Static spacetimes haunted by a phantom scalar field. II. Dilatonic charged solutions. \emph{Phys. Rev. D} \textbf{2021}, \emph{103}, {024004}.

\bibitem{sim:2019} Simpson, A.; Visser, M. Black-bounce to traversable wormhole. \emph{J. Cosmol. Astropart. Phys. } \textbf{2019}, {\emph{02}}, 042.

\bibitem{nojiri:2024} Nojiri, S.; Odintsov, S.D.; Folomeev, V. Wormholes inside stars and black holes. \emph{Phys. Rev. D} \textbf{2024}, \emph{109}, {104007}.

\bibitem{mann:1997} Mann, R.B. Black Holes of Negative Mass. \emph{Class. Quant. Gravit.} \textbf{1997}, {\emph{14}}, 2927--2930.

\bibitem{hull:2023}Hull, B.R.; Mann, R.B. Negative mass black holes in de Sitter space. \emph{Phys. Rev. D} \textbf{2023}, {\emph{107}}, 064027.

\bibitem{noz:2021} Nozawa, M. Static spacetimes haunted by a phantom scalar field. III. Asymptotically (A)dS solutions. \emph{Phys. Rev. D} \textbf{2021}, \mbox{\emph{103}, {024005}.}

\end{thebibliography}
\end{document}